\documentclass[12pt]{article}
\usepackage{amssymb}
\usepackage{graphicx}
\usepackage{makeidx}
\usepackage[dvips]{epsfig}
\usepackage{multicol}

\newcommand\cpc[3]   {
		{{\it Comput.\ Phys.\ Commun.\ }{\bf #1} (#2) #3}}
\newcommand\epjc[3]  {
		{{\it Eur.\ Phys.\ J. }{\bf C #1} (#2) #3}}
\newcommand\jhep[3]  {
		{{\it J. High Energy Phys.\ }{\bf #1} (#2) #3}}
\newcommand\nim[3]   {
		{{\it Nucl.\ Instrum.\ Meth.\ }{\bf #1} (#2) #3}}

\newcommand\npb[3]    {
		{{\it Nucl.\ Phys.\ }{\bf B #1} (#2) #3}}

\newcommand\plb[3]   {
		{{\it Phys.\ Lett.\ }{\bf B #1} (#2) #3}}

\newcommand\prl[3]   {
		{{\it Phys.\ Rev.\ Lett.\ }{\bf #1} (#2) #3}}
\newcommand\ptp[3]   {
		{{\it Prog.\ Theor.\ Phys.\ }{\bf #1} (#2) #3}}
\newcommand\ptps[3]   {
		{{\it Prog.\ Theor.\ Phys.\ Suppl.\ }{\bf #1} (#2) #3}}

\def \etal     {\relax\ifmmode{et \; al.}\else{$et \; al.$}\fi}
\def \hepph    {hep-ph/}

\begin{document}

\title{GR@PPA 2.7 event generator for $pp$/$p\bar{p}$ collisions}

\author{
S. Tsuno\footnote{Corresponding author, e-mail : Soushi.Tsuno@cern.ch} \\
{\itshape Department of Physics Faculty of Science, Okayama University} \\
{\itshape 3-1-1 Tsushima-naka, Okayama 700-8530, Japan} \\
{} \\
T. Kaneko, Y. Kurihara, and S. Odaka \\
{\itshape High Energy Accelerator Research Organization(KEK)} \\
{\itshape Tsukuba, Ibaraki 305-0801, Japan} \\
{} \\
K. Kato \\
{\itshape Kogakuin University} \\
{\itshape Nishi-Shinjuku 1-24, Shinjuku, Tokyo 163-8677, Japan} \\
}

\maketitle

\vspace*{-0.8cm}
\begin{abstract}
The GR@PPA event generator has been updated to version 2.7. This distribution 
provides event generators for $V$ ($W$ or $Z$) + jets ($\leq$ 4 jets), $VV$ + 
jets ($\leq$ 2 jets) and QCD multi-jet ($\leq$ 4 jets) production processes at 
$pp$ and $p\bar{p}$ collisions, in addition to the four bottom quark 
productions implemented in our previous work (GR@PPA\_4b). Also included are 
the top-pair and top-pair + jet production processes, where the correlation 
between the decay products are fully reproduced at the tree level. Namely, 
processes up to seven-body productions can be simulated, based on ordinary 
Feynman diagram calculations at the tree level. In this version, the GR@PPA 
framework and the process dependent matrix-element routines are separately 
provided. This makes it easier to add further new processes, and allows users 
to make a choice of processes to implement. This version also has several new 
features to handle complicated multi-body production processes. A systematic 
way to combine many subprocesses to a single base-subprocess has been 
introduced, and a new method has been adopted to calculate the color factors 
of complicated QCD processes. They speed up the calculation significantly.
\end{abstract}

\section{Introduction}

The great success of the Standard Model in recent decades leaves no doubt that 
gauge theories are capable of describing the interactions between elementary 
particles. Precise perturbative calculations based on gauge theories will 
become a crucial portion as larger collision energies are available to probe 
higher energy phenomena. For instance, the Higgs boson(s) and new particles 
predicted by theories beyond the Standard Model are the most important 
subjects to study in the current and future hadron collider experiments. They 
frequently decay to final states including multiple hadron jets, which 
predominant QCD interactions can easily imitate. A precise evaluation of these 
backgrounds is, therefore, necessary in order to accomplish a reliable 
measurement. However, once we try to do that using perturbative gauge 
theories, we immediately encounter a huge number of processes to evaluate. It 
is often too large to calculate by hand.

We have been carrying on automatic computations of Feynman diagrams by using 
the GRACE system \cite{grace}. GRACE is capable of calculations at the 
one-loop level \cite{grcloop} as well as at the tree level in the standard 
electroweak theory, and at the tree level in the minimal supersymmetric 
extension of the Standard Model (MSSM) \cite{grcmssm}. It can also be applied 
to QCD interactions. However, since the development has been mostly aimed at 
applications to lepton collisions, it is not directly applicable to 
hadron-collision interactions, mainly due to the necessity to include the 
parton distribution function (PDF). Besides, processes in hadron collisions 
usually consist of lots of subprocesses. They are desired to be combined into 
a single process.

In order to implement those features specific to hadron collisions, we have 
developed an extended framework, called GR@PPA (GRace At PP/Anti-p). Early 
developments can be seen in our previous reports \cite{abe,odaka}. The primary 
function of GR@PPA is to determine the initial and final state partons, 
$i.e.$, their flavors and momenta of the incoming partons by referring to a 
PDF, and those of the final state partons for jets or decay products. Based on 
the GRACE output codes, GR@PPA calculates the cross section and generates 
unweighted parton-level events using BASES/SPRING \cite{bases}. The GR@PPA 
framework also includes an interface to the LHA common data format 
\cite{leshouches}. The final steps of the event generation, $i.e.$, the 
initial- and final-state radiation, hadronization, decay and so forth, are 
done by passing the unweighted partonic events to PYTHIA \cite{pythia} or 
HERWIG \cite{herwig}.

Although the GR@PPA framework is not process-specific, we provide it as an 
event generator package including matrix elements for some selected processes. 
At the moment (version 2.7), the included processes are the weak boson(s) plus 
$N$ jets and $t\bar{t}$ plus $N$ jets productions as well as the pure QCD $N$ 
jets productions, where $N$ $\leq$ 4. These processes are the most important 
backgrounds in the Higgs boson and SUSY particle studies, and also useful to 
understand multi-body particle dynamics in precise measurements. Our previous 
work, the four bottom quark production (GR@PPA\_4b \cite{grappa_4b}), is also 
included. 

The reasons why we provide particular processes apart from the benefit of the 
automatic calculation by GRACE are the following; First, kinematical 
singularities in each process must be cared with a proper treatment. One can 
get a high efficiency to generate unweighted events since the kinematics are 
already well optimized. This is immediately addressed to the program running 
speed. It is critical for large scale MC productions. Second, it will be 
easier to adopt higher order calculations. Once the customized matrix element 
for an NLO process is prepared \cite{kuriharanlo}, users will be able to 
simply use it without any detailed care. Third, some different extensions are 
possible only by the modification of the framework. For example, GR@PPA 
generators can be extended to work under the C++ environment just by 
rewriting the framework in C++. The parton shower algorithm, $e.g.$ the NLL 
parton shower \cite{nll}, will also be possible to implement by modifying the 
framework.

We describe a symbolic treatment of the parton flavor in the diagram 
calculation in the next section. The feature of GR@PPA and its running are 
given in section 3. Some benchmark cross sections and program performances are 
presented in Sections 4 and 5, respectively. Finally, a summary is given in 
Section 6. We provide a list of supported processes and a summary of benchmark 
test conditions in Appendices.

\section{New features in the extension of GRACE to $pp$/$p\bar{p}$ collisions}

In hadron-hadron collisions, a certain process of interest usually consists of 
several incoherent subprocesses. The total cross section is thus expressed as 
a summation of the subprocesses as 
\begin{equation}
\sigma = \sum_{i, j, F} \int dx_{1} \int dx_{2} \int d\hat{\Phi}_{F} 
f^{1}_{i}(x_{1},Q^{2}) f^{2}_{j}(x_{2},Q^{2}) 
{ d\hat{\sigma}_{i j \rightarrow F}(\hat{s}) \over d\hat{\Phi}_{F} },
\label{eq:xsec}
\end{equation}
where $f^{a}_{i}(x_{a},Q^{2})$ is the PDF of the hadron 
$a$ ($p$ or $\bar{p}$), which gives the probability to find the parton $i$ 
with the energy fraction $x_{a}$ at the probing virtuality of $Q^{2}$. The 
differential cross section 
$d\hat{\sigma}_{i j \rightarrow F}(\hat{s})/d\hat{\Phi}_{F}$ describes the 
parton-level hard interaction producing the final-state $F$ from the collision 
of partons, $i$ and $j$, where $\hat{s}$ is the square of the total initial 
4-momentum. The sum is taken over all relevant combinations of $i$, $j$ and 
$F$. 

The original GRACE system assumes that both the initial and final states are 
well-defined. Hence, it can be applied to evaluating 
$d\hat{\sigma}_{i j \rightarrow F}(\hat{s})/d\hat{\Phi}_{F}$ and its 
integration over the final-state phase space $\hat{\Phi}_{F}$ only. An 
adequate extension is necessary to take into account the variation of the 
initial and final states. We have made two sorts of development in GR@PPA, 
--- applying PDF in the phase space integration and sharing several 
subprocesses as a single base-subprocess. The former is described in our 
previous paper \cite{grappa_4b}. Here, we focus on the later. As already 
mentioned, a "process" of interest is usually composed of several incoherent 
subprocesses in hadron interactions. In many cases, the difference between the 
subprocesses is only in the quark combination in the initial and/or final 
states. The matrix element of these subprocesses is frequently identical, or 
the difference is only in a few coupling parameters and/or masses. In such 
cases, it is convenient to add one more integration/differentiation variable 
to replace the summation in Eq. (\ref{eq:xsec}) with an integration. As a 
result, these subprocesses can share an identical "GRACE output code" and can 
be treated as a single base-subprocess. This modification drastically saves 
the program running time and simplifies the program coding.

The number of combinations to take $N$ out of $M$ flavors, allowing an 
overlap, is given by $_{M}H_{N}$($\equiv$ $\frac{(N+M-1)!}{N!(M-1)!}$). In 
case that all parton flavors up to the $b$-quark are considered, $M$ is equal 
to 11$(u,d,c,s,b,g,\bar{u},\bar{d},\bar{c},\bar{s},\bar{b})$. The 
configuration of the $N$ jets final state has $_{11}H_{N}$ subprocesses. 
Clearly a smaller $M$ decreases the number of subprocesses. Since the diagram 
structure is invariant against the quark (anti-quark) flavor exchange in QCD, 
the invariance is preserved by introducing generic up-type and down-type 
quarks even if the electroweak interaction is included. The base-subprocesses 
can be configured using these generic quarks. Then, the number of 
configurations is reduced to $_{5}H_{N}$ $\ll$ $_{11}H_{N}$. The output code 
of the matrix element from the GRACE has been so modified that the masses and 
couplings can be treated as input variables. The number of combinations has 
been further reduced using symmetries in the Standard Model such as the 
invariance in the charge and parity conjugate. In Table \ref{tab:subproc} we 
list up the number of base-subprocesses for the $N$ jets production process in 
$pp(\bar{p})$ collisions, to be compared with the total number of subprocesses 
that they cover. The initial colliding partons in the base-subprocesses in 
Table \ref{tab:subproc} are composed of $q_{u}q_{u}(\bar{q_{d}}\bar{q_{d}})$, 
$q_{u}\bar{q_{d}}$, $q_{u}g(q_{d}g)$, $q_{u}q_{d}$, 
$q_{u}\bar{q_{u}}(q_{d}\bar{q_{d}})$ and $gg$, where $q_{u}$($q_{d}$) and $g$ 
are the up(down)-type quark, and the gluon, respectively. 

Equation (\ref{eq:xsec2}) can be rewritten in term of the base-subprocess 
using a weight factor to treat the initial and final state parton 
configuration as 
\begin{equation}
\sigma = \sum_{i,j,F} \int dx_{1} \int dx_{2} \int d\hat{\Phi}_{F} \ 
w_{ijF} \ \frac{d\hat{\sigma}_{i j \rightarrow F}^{\mathrm{selected}}
(\hat{s};m,\alpha)}{d\hat{\Phi}_{F}} ,
\label{eq:xsec2}
\end{equation}
where $d\hat{\sigma}_{i j \rightarrow F}^{\mathrm{selected}}$ is the 
differential cross section of the base-subprocess with input arguments of 
masses and couplings. The coefficient $w_{ijF}$ is the weight factor for the 
chosen configuration. An appropriate subprocess is selected in event by event 
based on the flavor configuration decided by the weight factor. Since the QCD 
interactions are identical for any configuration in the subprocess, the weight 
factor is composed of PDF and the CKM (Cabibbo-Kobayashi-Maskawa) \cite{ckm} 
matrix as 
\begin{equation}
w_{ijF} = f^{1}_{i}(x_{1},Q^{2}) f^{2}_{j}(x_{2},Q^{2}) |V_{\mathrm{CKM}}|^{2K}
\{\mathrm{Br}(X \rightarrow F') \times \Gamma_{\mathrm{tot}}^{X}\}^{L} \; ,
\label{eq:xsec3}
\end{equation}
where $L$ is the number of $X$ bosons ($W$ or $Z$). The probability of the 
decay final state $F'$ can be handled in terms of the branching ratio, if the 
interference with the other partons is ignored. The branching ratio and the 
total width can be given by the experimentally measured ones. The flavor 
configuration is determined by the $|V_{\mathrm{CKM}}|^{2K}$, where $K$ is the 
number of $W$ bosons in the process.

We have to define a certain QCD color base representation in order to 
calculate the color factor of the interaction. In addition, PS programs 
usually require the information of color connection between partons in the 
event. The color bases should be so defined that the required connection 
information can be easily derived. In our previous works \cite{grc4f} we 
adopted such a definition that all color bases are composed of the SU(3) 
triplets. Namely, gluons are decomposed to a pair of quark and anti-quark. 
This definition includes color-singlet components which are not well 
interpreted for the hadronization. Besides, this leads to a large number of 
color bases for complicated processes consisting of many partons.

In the present version of GR@PPA we have adopted a different definition. We 
use a chain of a quark, $n$ gluons ($n$ = 0, 1, 2, ...) and an anti-quark as 
fundamental components. The color bases are given by products of such chains, 
and "glueballs" having no quark at the ends. This definition fits to the 
requirement of the LHA event interface \cite{leshouches}. In the event 
generation, one of the possible color bases is chosen in proportion to their 
squared amplitudes in order to determine the color connection. The 
interference is ignored ({\it i.e.}, a large $N_{c}$ approximation) as usual, 
since no reasonable way is known to include it. Of course, the interference is 
taken into account in the cross section evaluation. Compared to our previous 
definition, this definition significantly reduces the number of color bases for
complicated processes. For instance, it is reduced from 24 to 
14\footnote{Among 14 combinations, two color factors produce zero amplitudes 
at tree level.} for processes having two quark pairs and two gluons 
({\it e.g.}, $gg$ $\rightarrow$ $q\bar{q}q\bar{q}$). This reduction results in 
a faster calculation of the color factor and the color connection.


\begin{table}[htbp]
\begin{center}
\begin{tabular}{c|c|c} \hline
$N$ jets ($\alpha_{s}^{N}$) & base-subproc. & subproc. w/ all flavors \\ \hline
  2                         & 8             & 176                     \\
  3                         & 9             & 276                     \\
  4                         & 14            & 891                     \\ \hline
\end{tabular}
\end{center}
\caption{Number of base-subprocesses for the $N$ jet production process in 
$pp(\bar{p})$ collisions to be compared with the total number of subprocesses 
that they cover. The subprocesses can be classified according to the 
difference in the initial-state parton combination, and further classified by 
accounting for the charge and parity symmetries in the jet flavors.}
\label{tab:subproc}
\end{table}

\section{Program running}

\subsection{Distribution package}

The distribution package of the GR@PPA event generator consists of two part, 
the framework and the matrix elements. The framework contains a CHANEL 
library, a BASES/ SPRING package, and a kinematics library. The CHANEL 
library is a set of subroutines to calculate diagram elements: vertices, 
propagators and external legs. BASES/SPRING is a multi-dimensional 
general-purpose Monte Carlo integration and event-generation program set. The 
kinematics library converts the random numbers given by BASES/SPRING to a set 
of kinematical variables for the matrix element calculation, and converts the 
returned matrix element to the differential cross section. These routines are 
common to all processes. On the other hand, the matrix elements, generated by 
GRACE with some modifications for hadron collisions, are process dependent. 
They are supplied as separate packages.

The program running of GR@PPA generator is totally controlled by the 
subroutine {\tt GRCPYGEN}. In the initialization, {\tt GRCPYGEN} calls BASES 
to evaluate the total cross section for the given process, while it calls 
SPRING in the event generation cycle. The calling sequence of {\tt GRCPYGEN} 
is as follows: \\

{\tt CALL GRCPYGEN(CBEAM, IGSUB, MODE, SIGMA)},

\ \\
where the input arguments are

\begin{tabbing}
{\tt CBEAM (CHARACTER)} \= : '{\tt PP}' for $pp$ collisions 
and '{\tt PAP}' for $p\bar{p}$ collisions \\
{\tt IGSUB (INTEGER)} \> : process number \\
{\tt MODE (INTEGER)} 
\> : = 1 for calling BASES, and 0 for calling SPRING,
\end{tabbing}
and the output is \\

{\tt SIGMA (REAL*8)} : integrated cross section.

\ \\
The argument {\tt CBEAM} is dummy when {\tt MODE} = 0. A unique number 
{\tt IGSUB} is assigned to every physics process as listed in Appendix A. The 
output {\tt SIGMA} is always equal to the integrated cross section of the 
process specified by {\tt IGSUB}. The generated event is stored in the LHA 
common block \cite{leshouches}.

The distribution package is arranged for the use on Unix systems. However, 
since the structure is rather simple, we expect that the program can be 
compiled and executed on other platforms without serious difficulties. The 
package for the GR@PPA framework is composed of the following files and 
directories:

\begin{tabbing}
{\tt Config.perl} $\quad$ \= : \= a script to configure the setup, \\
{\tt Makefile} \> : \> {\tt Makefile} for the setup,\\
{\tt README} \> : \> a file describing how to set up the programs,\\
{\tt VERSION-2.76} \> : \> a note for this version,\\
{\tt proc.list} \> : \> a list of the processes supported in this version,\\
{\tt basesv5.1/} \> : \> BASES/SPRING (version 5.1) source codes,\\
{\tt chanel/} \> : \> CHANEL source codes,\\
{\tt grckinem/} \> : \> source codes of kinematics,\\
{\tt example/} \> : \> source codes of example programs,\\
{\tt inc/} \> : \> {\tt INCLUDE} files,\\
{\tt lib/} \> : \> the directory to store object libraries; initially empty.
\end{tabbing}

The {\tt proc.list} contains all process names and the corresponding process 
identification numbers {\tt IGSUB}. It also includes keywords for the 
processes used to configure the matrix elements.

The matrix element packages are separately given. They need to be installed in 
the GR@PPA framework directory. We provide several packages. Users can choose 
any of them according to their needs. Each matrix element package is composed 
of the following directories:

\begin{tabbing}
{\tt xxx} $\qquad$ \= : \= a set of matrix elements, \\
{\tt diagram} \> : \> Feynman diagrams used in the calculation of this process,
\end{tabbing}
where {\tt xxx} is a process group name. For example, the matrix elements for 
top quark production processes (with or without extra jets) are all placed 
under the {\tt top} directory.

\subsection{How to install}

The programs can be downloaded from \\

{\tt URL: http://atlas.kek.jp/physics/nlo-wg/grappa.html} . \\

The first task for the setup is to run the configuration script 
{\tt Config.perl}. Next, users have to edit the file {\tt Makefile} to specify 
the paths to the libraries to be linked, such as PYTHIA, HERWIG and CERNLIB. 
Those parts to be edited can be found at the top of the {\tt Makefile}. Users 
may also edit the {\tt inc/define.h} file to select a PDF library. Currently, 
PDFLIB in CERNLIB \cite{pdflib}, LHAPDF \cite{lhapdf}, and PYTHIA built-in 
PDFs are supported. The standalone CTEQ6 \cite{cteqpdf} routine is also 
included for the default use. In the case for using PDFLIB, edit one of the 
lines in {\tt inc/define.h} as \\

{\tt \#define PDFLIB  CERNPDFLIB} . \\

As an example we show an instruction to install the $W$+1jet and $W$+2jets 
production processes in the GR@PPA framework. After downloading the framework 
package and the two matrix element packages, unpack them as
\renewcommand{\baselinestretch}{0.78}
\begin{verbatim}
   tar xzvf GR@PPA-2.76.tgz
   cd GR@PPA-2.76
   tar xzvf ../matrix_w1j_v1.03.tgz
   tar xzvf ../matrix_w2j_v1.03.tgz .
\end{verbatim}
By these commands, the {\tt wjets} directory is created and under it two 
directories, {\tt w1j} and {\tt w2j}, are created. Then, configure the matrix 
elements and the compiling environment in 
{\tt Config.perl}, and execute it as 
\renewcommand{\baselinestretch}{0.78}
\begin{verbatim}
   Config.perl .
\end{verbatim}
This creates the {\tt Makefile} according to the present matrix-element 
installation. Users have to edit the {\tt Makefile} to specify further 
details, such as particular compiler options and paths to external libraries. 
Now, users can create the matrix element libraries and the framework 
libraries. Type the commands as follows:
\renewcommand{\baselinestretch}{0.78}
\begin{verbatim}
   make w1j
   make w2j
   make kinem
   make integ .
\end{verbatim}
Note that, because the kinematics codes are modified according to the selected 
matrix elements, the kinematics library must be created after making the 
matrix element libraries. Users must repeat the above sequence again when they 
add new matrix elements. The created libraries are installed in the {\tt lib/} 
directory by executing the command, 
\renewcommand{\baselinestretch}{0.78}
\begin{verbatim}
   make install .
\end{verbatim}
The example programs are given for the standalone use, and PYTHIA- and 
HERWIG-interfacing. The command 
\renewcommand{\baselinestretch}{0.78}
\begin{verbatim}
   make example
\end{verbatim}
sets up these examples in the {\tt example/} directory.

\subsection{Initialization and customization}

Although the execution of GR@PPA is controlled by the subroutine 
{\tt GRCPYGEN}, the detailed behavior depends on some parameters in common 
blocks and conditions defined in some subprograms. The subroutine 
{\tt GRCINIT} initializes all related parameters to the default values. Users 
can change them after calling {\tt GRCINIT} as described in the following.

The parameter that is necessary to be given by users is {\tt GRCECM}, which 
specifies the cm energy of the beam collision in GeV. Optionally, users can 
define some phase-space cuts in the laboratory frame: {\tt GPTCUT}, 
{\tt GETACUT} and {\tt GRCONCUT}. These parameters define the minimum $p_{T}$ 
in GeV, the largest pseudorapidity in the absolute value and the minimum 
separation in $\Delta R$, respectively. These cuts are applied to all produced 
jets except for those from weak boson decays. The separation ($\Delta R$) is 
defined for every pair of jets as 
\begin{equation}
\Delta R = \sqrt{\Delta\phi^{2} + \Delta\eta^{2}},
\label{eq:DeltaR}
\end{equation}
where $\Delta\phi$ and $\Delta\eta$ are the separation in the azimuthal angle 
and the pseudorapidity, respectively. The arrays {\tt GRCPTCUT}, 
{\tt GRCETACUT} and {\tt GRCRCONCUT} can separately set certain phase-space 
cuts for each final state particle.

Additionally, the subroutine {\tt GRCUSRCUT} provides a framework to apply 
customized cuts referring to detailed properties such as four-momenta of the 
particles. Some useful functions are available there. If the events are 
acceptable {\tt IUSRCUT = 0} should be returned, while {\tt IUSRCUT = 1} for 
the rejection. Since those cuts are applied at the integration stage, a high 
event generation efficiency can be achieved in the event generation. The 
selection can be applied even for the particle flavor. This feature is useful 
for a parton flavor selection in the jet production. A concrete example can be 
found in Appendix B.

The arrays {\tt IGWMOD} and {\tt IGZMOD} specify the decay mode of the $W$ and 
$Z$ bosons, respectively. If the value is equal to 1, the corresponding decay 
channel is activated; if the number is equal to 0, the channel is disabled. 
{\tt IWIDCOR} is an option for the decay width correction for the $W$ and $Z$ 
bosons. The theoretically calculated values are used if {\tt IWIDCOR = 1}, 
while the experimentally measured ones can be used if {\tt IWIDCOR = 2}. If 
{\tt IWIDCOR = 2}, the cross section is weighted by using the measured 
branching ratio as
\begin{equation}
\sigma(X \rightarrow f\bar{f}) \; = \; \sigma_{\mathrm{calc}} \
\frac{\Gamma_{\mathrm{exp}}^{\mathrm{tot}}}
{\Gamma_{\mathrm{calc}}^{\mathrm{part}}} \ 
\mathrm{Br}_{\mathrm{exp}}(X \rightarrow f\bar{f}) \quad ,
\label{eq:xseccalc}
\end{equation}
where $\sigma_{\mathrm{calc}}$ is the calculated cross section using the given 
total decay width $\Gamma_{\mathrm{exp}}^{\mathrm{tot}}$, 
$\Gamma_{\mathrm{calc}}^{\mathrm{part}}$ is the partial decay width 
theoretically calculated, and 
$\mathrm{Br}_{\mathrm{exp}}(X \rightarrow f\bar{f})$ is the branching ratio 
experimentally measured. $\mathrm{Br}_{\mathrm{exp}}(X \rightarrow f\bar{f})$ 
must be given in the arrays {\tt GRCWBR} and {\tt GRCZBR} for the $W$ and $Z$ 
bosons, respectively. 

The array {\tt IGJFLV} determines the active flavor of jets in the final state 
except for those from $W$ and $Z$ boson decays. If {\tt IGJFLV(I) = 1}, the 
corresponding flavor $I$ ($d,u,s,c,b,t,g$ for $I=1,2,...,7$) is activated, 
while it is disabled if {\tt IGJFLV(I) = 0}. Regardless that they are disabled 
by the {\tt IGJFLV} flags, those flavors are occasionally produced if the 
diagram structure requires them. For example, consider the case that the $u$ 
flavor in the PDF is activated and the $qg$ $\rightarrow$ $qg$ base-subprocess 
is chosen in the QCD 2 jets production. In this case, if $g$ is activated, the 
$ug$ $\rightarrow$ $ug$ subprocess is activated even if $u$ is disabled by 
{\tt IGJFLV}. If both $u$ and $g$ are disabled, this subprocess itself is 
disabled. That is, the events are accepted if either of the activated flavors 
by {\tt IGJFLV} has to appear in the final state. The jet flavor can also be 
controlled by changing the CKM parameter {\tt GRCCKM} if the process includes 
intermediate $W$ bosons. 

Further detailed parameters, such as particle masses, decay widths and 
couplings, are accessible in the subroutine {\tt SETMAS}. The mass and the 
total decay width of the weak bosons can be manually controlled there. GR@PPA 
does not give any constraint to these parameters. However, the electroweak 
parameters are characterized by only three (plus Higgs mass) 
parameters.\footnote{Additionally, a gauge invariance also demands fermion 
masses in the Yukawa sector.} The setting for dependent parameters are ignored 
and their values are recalculated according to the choice of the scheme given 
by the parameter {\tt IGAUGE}. The $G_{\mu}$ scheme is taken as the default, 
where all the parameters are given by the set of following parameters:
\begin{equation}
 (G_{F}, \; M_{W}, \; M_{Z}) \; .
\label{eq:gaugeinv}
\end{equation}
Here, $G_{F}$ is the Fermi constant, $M_{W}$ and $M_{Z}$ are masses of $W$ and 
$Z$ bosons, respectively.

The other process-specific parameters are defined in the subroutine 
{\tt GRCPAR}. Users can choose different conditions for different processes. 
The variable {\tt ICOUP} determines the energy scale ($Q^{2}$) for calculating 
the coupling strengths, $\alpha_{\mathrm{em}}$ and $\alpha_{s}$, in the 
matrix element calculation (renormalization scale). The selectable choices are 
prepared (see Table \ref{tab:parameter}). The variable {\tt IFACT} determines 
$Q^{2}$ for PDF (factorization scale). The definition is the same as 
{\tt ICOUP}. The same choice as {\tt ICOUP} is taken if {\tt IFACT} is not 
explicitly given. As an option, users can apply their own definitions of these 
energy scales, by setting {\tt ICOUP = 6} and/or {\tt IFACT = 6} and editing 
the subroutine {\tt GRCUSRSETQ}. An example is attached to {\tt grcpar.F}. The 
number of flavors used in the coupling calculation and PDF is given in 
{\tt INPFL}. The flavor is ordered as $u$, $d$, $c$, $s$ and $b$. The 
extrapolation to the top quark in the coupling formula and PDF is not taken 
into account at the moment.

The integer variable {\tt IBSWRT} controls whether BASES should be called in 
the initialization or not. The task of BASES is to optimize the integration 
grids and, after that, store the optimized results in a ``BASES table''. The 
execution of BASES consumes much CPU time because a precise evaluation is 
necessary for an efficient event generation by SPRING. It is not necessary to 
repeat the execution for identical conditions. A previously optimized result 
(``BASES table'') is reused if {\tt IBSWRT = 1}. It should be noted that, once 
the parameters which affects the estimated result of the integration are 
changed, the "condition" is no longer identical and BASES has to be 
re-executed. The parameter {\tt NCALL} specifies the number of sampling points 
in each step of the iterative grid optimization in BASES. The larger this 
number is, the better the conversion would be. However, it takes longer in the 
CPU time. The optimized values are preset in {\tt grcpar.F}. The character 
variable {\tt GRCFILE} gives the ``BASES table'' file name\footnote{BASES 
actually creates two files having extensions of .data and .result, 
respectively, added to the name given by {\tt GRCFILE}. The former is the 
``BASES table'', while the latter is a readable summary of the BASES 
execution.}. A new file must be specified if {\tt IBSWRT = 0}, while an 
existing file must be specified if {\tt IBSWRT = 1}. 

The parameters described in this subsection are summarized in 
Table \ref{tab:parameter}. 

\begin{table}[htbp]
\begin{center}
\begin{tabular}{|l|l|} \hline
Parameter & Description \\ \hline \hline

{\tt GRCECM} & CM energy of the beam collision in GeV. (D=14000.) \\ \hline

{\tt CBEAM} & '{\tt PP}' for $pp$ collisions and '{\tt PAP}' for $p\bar{p}$ 
collisions. (D='PP')\\ \hline

{\tt IGSUB} & Process number. 
See Table \ref{tab:proc}. \\ \hline

{\tt MODE} & Mode selection for {\tt GRCPYGEN}: = 1 for calling BASES, 
and 0 for \\
& calling SPRING. \\ \hline

{\tt GPTCUT} & Minimum $p_{T}$ cut in GeV for jets except for those from 
weak \\
& boson decays. (D=20.GeV) \\ 
{\tt GRCPTCUT(8)} & Minimum $p_{T}$ cut in GeV for each final state particle.
(D=0.GeV) \\ \hline

{\tt GETACUT} & Largest pseudorapidity cut in the absolute value for jets 
except for \\
& those from weak boson decays.(D=3.) \\ 
{\tt GRCETACUT(8)} & Largest pseudorapidity cut for each final state particle.
(D=10.) \\ \hline

{\tt GRCONCUT} & Minimum separation cut in $\Delta R$ 
(= $\sqrt{\Delta\phi^{2}+\Delta\eta^{2}}$) for jets except \\
& for those from weak boson decays.(D=0.4) \\ 
{\tt GRCRCONCUT(8)} & Minimum separation cut in $\Delta R$ 
(= $\sqrt{\Delta\phi^{2}+\Delta\eta^{2}}$) for each final \\
& state particle. (D=0.) \\ \hline

{\tt IGWMOD(20)} & Decay mode for $W$ boson: = 1 to activate, and 0 to 
deactivate. \\ \hline

{\tt IGZMOD(16)} & Decay mode for $Z$ boson : = 1 to activate, and 0 to 
deactivate. \\ \hline

{\tt IWIDCOR} & Width correction for $W$ and $Z$. (D=1) \\ \hline

{\tt GRCWBR(20)} & Branching ratio for $W$. It is used only when 
{\tt IWIDCOR}=2. \\ \hline

{\tt GRCZBR(16)} & Branching ratio for $Z$. It is used only when 
{\tt IWIDCOR}=2. \\ \hline

{\tt IGJFLV(7)} & Flag for the jet flavor of jets except for those from weak 
boson \\
& decays: = 1 to activate, and 0 to deactivate. \\ \hline

{\tt IGRCGEF} & Photon interference in $Z$ production processes is included \\
& if set to 1, ignored if 0. (D=1) \\ \hline

{\tt GRCCKM(3,3)} & CKM parameters. \\ \hline

{\tt IGAUGE} & Choice of electroweak parameters. \\
& The $G_{\mu}$ scheme is used as the default. (D=1) \\ \hline

{\tt ICOUP} & Choice of the renormalization scale. \\
 & = 1 : $\sqrt{\hat{s}}$ of the hard interaction. \\
 & = 2 : average of squared transverse mass ($<m_{T}^{2}>$). \\
 & = 3 : total squared transverse mass ($\sum m_{T}^{2}$). \\
 & = 4 : maximum squared transverse mass ($max$ $m_{T}^{2}$). \\
 & = 5 : fixed value. Set {\tt GRCQ} in GeV. \\
 & = 6 : user defined scale. Set {\tt GRCQ} in the subroutine 
{\tt GRCUSRSETQ}. \\ \hline

{\tt IFACT} & Choice of the factorization scale. (D=0) \\
& The definition is the same as {\tt ICOUP}. \\
 & If {\tt IFACT} = 0, the same value as the renormalization scale is used. \\
 & In case {\tt IFACT} = 5 or 6, set {\tt GRCFAQ} in GeV. \\ \hline

{\tt GRCFILE} & Output file name for the BASES integration. \\ \hline

{\tt IBSWRT} & Mode selection for BASES integration: = 0 for calling the 
BASES \\

& integration, and 1 for skipping. (D=0) \\ 
& If {\tt IBSWRT} = 1, the file defined in {\tt GRCFILE} is used. \\ \hline

{\tt NCALL} & Number of sampling points in each step of the iterative grid \\

& optimization in BASES. \\ \hline

{\tt INPFL} & Number of flavors used in the coupling calculation and PDF.
(D=5) \\ \hline

\end{tabular}
\caption{Parameters in GR@PPA.}
\label{tab:parameter}
\end{center}
\end{table}

\subsection{BASES integration}

In the output of GR@PPA, users should pay appropriate attention to the print 
out from BASES, especially when they apply tight cuts. Since each subprocess 
is composed of many coherent diagrams, it is not practicable to take all 
singularities into account in the ``kinematics'' definitions. Some very minor 
ones are ignored in GR@PPA. A combination of very tight cuts may enhance the 
relative contribution of ignored singularities. In such cases, it is likely to 
happen that, in the BASES iteration, the estimated total cross section jumps 
(increases) to a value unreasonably different from the previous estimation 
and, accordingly, the estimated error increases. Users should consider that 
they must be in such a trouble if they find a jump of, for instance, more than 
three times the previous error. The results are unreliable in the phase-space 
region defined by such cuts. The instructive integration accuracy is $0.5$\% 
or better for every iteration. Users should change the parameter {\tt NCALL} 
to a larger value if this accuracy is not achieved.

\section{Processes}

\subsection{$V$ + $N$ jets}

The process identification number {\tt IGSUB} is assigned to be 100 $\sim$ 104 
and 110 $\sim$ 114 for $W$ and $Z$ production processes with 0 up to 4 jets. 
The matrix elements (ME) include the boson decays into two fermions, so that 
the decay properties are correctly reproduced at the tree level. Heavy quarks 
such as bottom and top quarks can also be included as the jets as well as the 
lighter quarks and the gluon. Since the quark mixing in the couplings with $W$ 
boson is treated up to the third generation, single and triple heavy quark 
productions are allowed in the multi-jet configuration, in addition to the 
pair production from the gluon splitting. In $Z$ + $N$ jets processes, the 
$Z$/$\gamma^{*}$ interference can be taken into account in the ME 
calculations. The interference between decay fermions and those from other 
sources is ignored.

The benchmark cross sections of the $V$ + $N$ jets production processes for 
the Tevatron Run II and the LHC conditions are presented in 
Table \ref{xsecvjet}, where the quark flavors are included up to the $b$ 
quark. We also present the benchmark cross sections for the case having at 
least one $b$ quark, and having a $t\bar{t}$ pair in the final state. The 
results are in good agreement with those from other 
generators \cite{alpgen,madgraph,comphep,amegic,phegas}. The detailed 
parameters used in these calculations are described in Appendix B.

\begin{table}[htbp]
\begin{center}
\begin{tabular}{c|c|c|c|c} \hline
  \multicolumn{1}{c}{\makebox[15mm]{}}
& \multicolumn{2}{|c}{\makebox[30mm]{Tevatron Run-II}}
& \multicolumn{2}{|c}{\makebox[30mm]{LHC}} \\ \hline

  \multicolumn{1}{c}{\makebox[15mm]{$N$ jets}}
& \multicolumn{1}{|c}{\makebox[15mm]{$W$($e\nu_{e}$)}}
& \multicolumn{1}{|c}{\makebox[15mm]{$Z$($e^{+}e^{-}$)}}
& \multicolumn{1}{|c}{\makebox[15mm]{$W$($e\nu_{e}$)}}
& \multicolumn{1}{|c}{\makebox[15mm]{$Z$($e^{+}e^{-}$)}} \\ \hline

0 & 1.576(2)$\times$10$^{3}$
  & 1.598(3)$\times$10$^{2}$
  & 1.116(2)$\times$10$^{4}$
  & 9.57(3)$\times$10$^{2}$ \\

1 & 1.852(3)$\times$10$^{2}$
  & 1.829(4)$\times$10$^{1}$
  & 2.854(5)$\times$10$^{3}$
  & 2.614(7)$\times$10$^{2}$ \\

2 & 3.461(7)$\times$10$^{1}$
  & 3.485(6)
  & 1.143(3)$\times$10$^{3}$
  & 1.082(2)$\times$10$^{2}$ \\

3 & 6.29(2)
  & 6.35(2)$\times$10$^{-1}$
  & 4.82(1)$\times$10$^{2}$ 
  & 4.53(1)$\times$10$^{1}$ \\

4 & 1.201(5) 
  & 1.173(3)$\times$10$^{-1}$
  & 2.19(1)$\times$10$^{2}$ 
  & 2.045(5)$\times$10$^{1}$ \\ \hline

2 ($\geq$ 1 $b$) & 3.260(6)$\times$10$^{-1}$
                 & 9.24(2)$\times$10$^{-2}$
                 & 2.720(7)
                 & 8.68(2) \\

3 ($\geq$ 1 $b$) & 1.019(2)$\times$10$^{-1}$
                 & 2.266(5)$\times$10$^{-2}$
                 & 4.305(9)
                 & 3.740(7) \\

4 ($\geq$ 1 $b$) & 2.947(8)$\times$10$^{-2}$
                 & 3.817(5)$\times$10$^{-3}$
                 & 3.90(3)
                 & 9.47(2)$\times$10$^{-1}$ \\ \hline

$t\bar{t}$ + 0 & 5.269(9)$\times$10$^{-4}$
               & 2.248(3)$\times$10$^{-4}$
               & 3.774(7)$\times$10$^{-2}$
               & 2.682(6)$\times$10$^{-2}$ \\

$t\bar{t}$ + 1 & 1.357(2)$\times$10$^{-4}$
               & 6.302(9)$\times$10$^{-5}$
               & 4.98(2)$\times$10$^{-2}$
               & 3.59(1)$\times$10$^{-2}$ \\

$t\bar{t}$ + 2 & 6.86(1)$\times$10$^{-5}$
               & 5.707(5)$\times$10$^{-7}$
               & 6.52(2)$\times$10$^{-2}$
               & 1.156(2)$\times$10$^{-2}$ \\ \hline
\end{tabular}
\end{center}
\caption{Benchmark cross section (pb) for $V$ + $N$ jets processes. Results 
are presented for the Tevatron Run II and the LHC cases. The detailed 
parameters used in the calculations are described in Appendix B.}
\label{xsecvjet}
\end{table}

\subsection{$VV$ + $N$ jets}

The number {\tt IGSUB} is assigned to be 120 $\sim$ 122 for the $WW$ 
production processes, 130 $\sim$ 132 for the $WZ$ production processes, and 
140 $\sim$ 142 for the $ZZ$ production processes. The matrix elements (ME) 
include the boson decays into two fermions, so that the decay properties such 
as the spin correlation between two bosons is correctly reproduced at the tree 
level, while the interference between the fermions from different bosons and 
those from other sources is ignored. The $Z$/$\gamma^{*}$ interference can be 
taken into account in the ME calculations. Heavy quarks ($b$ and $t$) can also 
be included in the jets.

The benchmark cross sections for the Tevatron Run II and the LHC conditions 
are presented in Table \ref{xsecvvjet}. We also present the results for the 
case having at least one $b$ quark, and having a $t\bar{t}$ pair in the final 
state. The detailed parameters are described in Appendix B.

\begin{table}[htbp]
\begin{center}
\begin{tabular}{c|c|c|c} \hline
  \multicolumn{1}{c}{\makebox[15mm]{}}
& \multicolumn{3}{|c}{\makebox[45mm]{Tevatron Run-II}}  \\ \hline

  \multicolumn{1}{c}{\makebox[15mm]{$N$ jets}}
& \multicolumn{1}{|c}{\makebox[15mm]{$WW$}}
& \multicolumn{1}{|c}{\makebox[15mm]{$WZ$}}
& \multicolumn{1}{|c}{\makebox[15mm]{$ZZ$}} \\ \hline
0 & 7.91(2)$\times$10$^{-2}$ 
  & 6.06(1)$\times$10$^{-3}$ 
  & 1.541(3)$\times$10$^{-3}$ \\

1 & 1.986(5)$\times$10$^{-2}$ 
  & 2.013(4)$\times$10$^{-3}$ 
  & 3.721(6)$\times$10$^{-4}$ \\

2 & 4.77(1)$\times$10$^{-3}$ 
  & 5.16(2)$\times$10$^{-4}$
  & 7.698(9)$\times$10$^{-5}$ \\ \hline

2 ($\geq$ 1 $b$) & 1.582(2)$\times$10$^{-4}$ 
                 & 6.00(2)$\times$10$^{-6}$ 
                 & 1.629(3)$\times$10$^{-6}$ \\ \hline

$t\bar{t}$ + 0 & 5.504(5)$\times$10$^{-7}$ 
               & 5.065(7)$\times$10$^{-8}$
               & 8.871(9)$\times$10$^{-9}$ \\ \hline
\end{tabular}
\end{center}

\begin{center}
\begin{tabular}{c|c|c|c} \hline
  \multicolumn{1}{c}{\makebox[15mm]{}}
& \multicolumn{3}{|c}{\makebox[45mm]{LHC}} \\ \hline

  \multicolumn{1}{c}{\makebox[15mm]{$N$ jets}}
& \multicolumn{1}{|c}{\makebox[15mm]{$WW$}}
& \multicolumn{1}{|c}{\makebox[15mm]{$WZ$}}
& \multicolumn{1}{|c}{\makebox[15mm]{$ZZ$}} \\ \hline
0 & 1.339(5)
  & 4.64(1)$\times$10$^{-2}$
  & 1.080(2)$\times$10$^{-2}$ \\

1 & 1.071(3)
  & 1.873(4)$\times$10$^{-1}$
  & 6.06(1)$\times$10$^{-3}$ \\

2 & 9.41(2)$\times$10$^{-1}$ 
  & 1.536(4)$\times$10$^{-1}$ 
  & 3.419(5)$\times$10$^{-3}$ \\ \hline

2 ($\geq$ 1 $b$) & 7.36(1)$\times$10$^{-2}$ 
                 & 1.164(3)$\times$10$^{-4}$
                 & 1.887(8)$\times$10$^{-4}$ \\ \hline

$t\bar{t}$ + 0 & 7.432(9)$\times$10$^{-4}$ 
               & 1.213(2)$\times$10$^{-5}$
               & 3.040(5)$\times$10$^{-6}$ \\ \hline
\end{tabular}
\end{center}

\caption{Benchmark cross section (pb) for $VV$ + $N$ jets processes. Results 
are presented for the Tevatron Run II and the LHC cases. The detailed 
parameters used in the calculations are described in Appendix B.}
\label{xsecvvjet}
\end{table}

\subsection{$b\bar{b}b\bar{b}$}

We provide the $b\bar{b}b\bar{b}$ production processes which have been 
developed in our previous work \cite{grappa_4b} as a separate matrix element 
package. The process identification number is 160 for the Higgs production 
associated with bottom quarks ($Y_{b}^{2}\alpha_{s}^{2}$), 161 for the 
$Z$/$\gamma^{*}$ mediated process ($\alpha_{s}^{2}\alpha_{\mathrm{em}}^{2}$), 
162 for the pure QCD process ($\alpha_{s}^{4}$), 163 for the $HZ$ production 
($Y_{b}^{2}\alpha_{\mathrm{em}}^{2}$), and 164 for the $Z$ pair mediated 
process ($\alpha_{\mathrm{em}}^{4}$), where the SM Higgs boson is assumed for 
the Higgs production.

\subsection{$t\bar{t}$ + $N$ jets}

Aside from the QCD multi-jets processes, we provide top pair production 
processes. In these processes, the whole decay chain of the top quark is 
included in the diagram calculation, so that the spin correlation in the top 
decay is fully reproduced. In addition to the $t\bar{t}$ production 
(6-body ME), we also provide the $t\bar{t}$ + 1 jet (7-body ME) process. This 
process may give an insight for the 7-body kinematics including the top decay. 
The process number is assigned to be 170 for $t\bar{t}$, and 171 for 
$t\bar{t}$ + 1 jet. Possible anomalous couplings in the top decay and its 
production will be included in the ME calculation in a future version. They 
will be provided as an update set of the matrix elements.

\subsection{$N$ jets}

The QCD multi-jets processes are provided with the process identification 
numbers from 182 to 184 for 2, 3 and 4 jets, respectively. The electroweak 
interaction is ignored in these processes. Heavy flavors such as bottom and 
top quarks can be produced by appropriately setting {\tt IGJFLV}. 

The benchmark cross sections of the QCD $N$ jets processes are presented in 
Table \ref{xsecqcdjet}. We also present the results for heavy flavor 
production processes. These results are in good agreement with those from 
other generators \cite{alpgen,madgraph,comphep,amegic,phegas}. It should be 
noted that the result for the $b\bar{b}b\bar{b}$ production is the same as 
that for the process {\tt IGSUB = 162}. The detailed parameter setting is 
described in Appendix B.

\begin{table}[htbp]
\begin{center}
\begin{tabular}{c|c|c} \hline
  \multicolumn{1}{c}{\makebox[15mm]{}}
& \multicolumn{1}{|c}{\makebox[30mm]{Tevatron Run-II}}
& \multicolumn{1}{|c}{\makebox[30mm]{LHC}} \\ \hline

  \multicolumn{1}{c}{\makebox[15mm]{$N$ jets}}
& \multicolumn{1}{|c}{\makebox[15mm]{QCD jets ($\alpha_{s}^{N}$)}}
& \multicolumn{1}{|c}{\makebox[15mm]{QCD jets ($\alpha_{s}^{N}$)}} \\ \hline

2 & 1.853(5)$\times$10$^{7}$ 
  & 4.355(8)$\times$10$^{8}$ \\

3 & 6.88(3)$\times$10$^{5}$ 
  & 3.326(7)$\times$10$^{7}$ \\

4 & 8.82(3)$\times$10$^{4}$ 
  & 7.94(1)$\times$10$^{6}$ \\ \hline

$b\bar{b}b\bar{b}$ & 5.70(1)
                   & 8.42(2)$\times$10$^{2}$ \\

$b\bar{b}t\bar{t}$ & 7.46(2)$\times$10$^{-3}$
                   & 3.796(8) \\

$t\bar{t}t\bar{t}$ & 5.809(4)$\times$10$^{-6}$
                   & 2.322(5)$\times$10$^{-2}$ \\ \hline

\end{tabular}
\end{center}
\caption{Benchmark cross section (pb) for QCD multi-jets processes. Results 
are presented for the Tevatron Run-II and the LHC cases. The detailed 
parameters used in the calculations are described in Appendix B.}
\label{xsecqcdjet}
\end{table}

\section{Performance}

The computation performance of GR@PPA for the $W$ + $N$ jets processes in the 
Tevatron Run-II condition is summarized in Table \ref{tab:xsecspeed}. The 
tests have been performed using Intel Pentium 4 3.4 GHz CPU and two different 
Fortran compilers: a free software, g77 in gcc 2.96, and a commercial 
compiler, Intel Fortran Compiler version 8.0. The integration time and the 
generation speed are separately shown. Clearly, the commercial compiler is 
about 2.5 times faster than the free compiler. However, in both cases, the 
time is not intolerable for large scale Monte Carlo productions. The 
generation efficiencies of SPRING, shown in the table, are exceptionally good 
for this kind of complicated processes. 

The result for the $W$ + 4 jets is not shown in the table, since it consumes 
too long CPU time to run as a single process. A parallel computing is 
indispensable for practical uses. A parallel computing employing MPI 
(Message Passing Interface) is supported in BASES. The benchmark result in 
Table \ref{xsecvjet} has been obtained by using a PC farm having 12 Intel 
Pentium 4 3.2 GHz CPUs with Intel Fortran Compiler version 7.0. The 
integration time is 23 hours fully using these CPUs. The event generation 
speed is 0.11 events/sec with the generation efficiency of 0.1\%. Though 
SPRING does not support parallel computing, users can parallelize the event 
generation by hand by changing the starting random number seed. The parallel 
computing has also been applied to the $W$ + 3 jets production. The 
integration time has been reduced to 24 minutes from 5 hours for the 
single-process computing on the same platform. The scalability is quite good. 
Those users who want to apply the parallel computing should contact the 
author\footnote{Soushi.Tsuno@cern.ch}.


\begin{table}[htbp]
\begin{center}
\begin{tabular}{c|c|c|c|c} \hline
Process & Fortran  & Integration    & Event Generation & Efficiency \\
        & Compiler & time (H:M:Sec) & speed (events/sec) & (\%) \\ \hline
$W$($e\nu_{e}$) + 0 jet  & g77       & 00 : 00 : 05 &  43859 & 69.6 \\
                         & intel 8.0 & 00 : 00 : 02 & 101010 &      \\ \hline
$W$($e\nu_{e}$) + 1 jet  & g77       & 00 : 00 : 52 &  13927 & 19.9 \\
                         & intel 8.0 & 00 : 00 : 19 &  34364 &      \\ \hline
$W$($e\nu_{e}$) + 2 jets & g77       & 00 : 38 : 27 &    709 &  1.7 \\
                         & intel 8.0 & 00 : 13 : 52 &   1957 &      \\ \hline
$W$($e\nu_{e}$) + 3 jets & g77       & 14 : 03 : 46 &     18 &  0.3 \\
                         & intel 8.0 & 04 : 57 : 50 &     52 &      \\ \hline
\end{tabular}
\end{center}
\caption{Performance of GR@PPA for the $W$ + $N$ jets processes in the 
Tevatron Run-II condition. The tests have been performed using Intel Pentium 4 
3.4 GHz CPU and two different Fortran compilers: g77 version 2.96, and Intel 
Fortran Compiler version 8.0. The integration time and the generation speed 
are separately shown.}
\label{tab:xsecspeed}
\end{table}

\section{Summary}

The GR@PPA event generator has been updated to version 2.7. In this version, 
the extensions have been mostly applied to deal with the flavor configuration 
in the initial and final state by sharing several subprocesses into the single 
subprocess. This allows us to incorporate the variation in the initial and 
final states parton configurations such as the multi-jets production 
processes. A new method has been also adopted to calculate the color factors 
of complicated QCD processes. Those implementations are able to speed up the 
diagram calculation for multi-jet production processes significantly.

The distribution of this version is composed of two separate packages: the 
framework and the matrix element for the processes. The framework consists of 
some process-independent elements such as integration packages and kinematic 
libraries. The matrix element provides a set of the matrix elements used in 
the GR@PPA event generator. Currently, we provide the matrix element packages 
for $V$ ($W$ or $Z$) + jets ($\leq$ 4 jets), $VV$ + jets ($\leq$ 2 jets) 
and QCD multi-jet ($\leq$ 4 jets) production processes at the tree level for 
$pp$ and $p\bar{p}$ collisions, where the jet flavor includes up to top quark. 
The four bottom quark productions implemented in our previous work 
(GR@PPA\_4b) are also included. In addition, we provide the top-pair and 
top-pair + jet production processes, where the correlation between the decay 
products are fully reproduced at the tree level. Namely, processes up to 
seven-body productions can be simulated, based on ordinary Feynman diagram 
calculations.

The GR@PPA event generator takes advantages that further extensions are easily 
applicable in the GR@PPA framework. Once NLO processes or non-Standard Model 
processes are given as a matrix element package, users will be able to simply 
use them without any detailed care. An extended framework to include the 
parton shower is also possible. They may be available in a future release or 
by a release of the new matrix element packages. The GR@PPA event generator 
will be suitable not only for a large scale Monte Carlo production for high 
luminosity hadron collisions at Tevatron and LHC, but also for future NLO 
calculations to be composed of lots of subprocesses.

\section{Acknowledgements}

This study has been carried out in a collaboration between the Atlas-Japan 
group formed by Japanese members of the Atlas experiment and the Minami-Tateya 
numerical calculation group lead by Y. Shimizu. We would like to thank all 
members of both groups. Particularly, we are grateful to J. Fujimoto and 
T. Ishikawa for useful guidance and comments. This work was supported in part 
by Japan society of the promotion of science, Grant-in-Aid for Scientific 
Research 14340081 and 17540283. This work was also supported in part by the 
``International Research Group'' on ``Automatic Computational Particle 
Physics''(IRG ACPP) funded by CERN/Universities in France, KEK/MEXT in Japan 
and MSU/RAS/MFBR in Russia.

\appendix

\section{Processes in GR@PPA}

All physics processes are recognized by unique process numbers {\tt IGSUB} 
listed in Table \ref{tab:proc}. The second column in the table is 
the order of coupling constants for the process. The parameter $Y_{b}$ is the 
Yukawa coupling of the $b$ quark. The third column shows the command paramter 
for compiling the ME. The parameter name is the same as the directory name of 
the process. The forth column is the process descriptions. Each process 
consists of several base-subprocesses. The symbol ``$f$'' denotes the fermion 
from boson decays. The decay fermions except for those in the 
$b\bar{b}b\bar{b}$ processes do not interfere with other partons.

\begin{table}[htbp]
\begin{center}
\begin{tabular}{c|c|c|c} \hline
{\tt IGSUB} & Coupling & Command   & Process description \\ 
            & order    & parameter & \\ \hline
100 & $\alpha_{\mathrm{em}}^{2}$ & {\tt w0j} 
& $pp(\bar{p})$ $\rightarrow$ $W(2f)$ + 0 jet (+$X$) \\

101 & $\alpha_{\mathrm{em}}^{2}\alpha_{s}$ & {\tt w1j} 
& $pp(\bar{p})$ $\rightarrow$ $W(2f)$ + 1 jet (+$X$) \\

102 & $\alpha_{\mathrm{em}}^{2}\alpha_{s}^{2}$ & {\tt w2j} 
& $pp(\bar{p})$ $\rightarrow$ $W(2f)$ + 2 jets (+$X$) \\

103 & $\alpha_{\mathrm{em}}^{2}\alpha_{s}^{3}$ & {\tt w3j} 
& $pp(\bar{p})$ $\rightarrow$ $W(2f)$ + 3 jets (+$X$) \\

104 & $\alpha_{\mathrm{em}}^{2}\alpha_{s}^{4}$ & {\tt w4j} 
& $pp(\bar{p})$ $\rightarrow$ $W(2f)$ + 4 jets (+$X$) \\ \hline

110 & $\alpha_{\mathrm{em}}^{2}$ & {\tt z0j} 
& $pp(\bar{p})$ $\rightarrow$ $Z/\gamma^{*}(2f)$ + 0 jet (+$X$) \\

111 & $\alpha_{\mathrm{em}}^{2}\alpha_{s}$ & {\tt z1j} 
& $pp(\bar{p})$ $\rightarrow$ $Z/\gamma^{*}(2f)$ + 1 jet (+$X$) \\

112 & $\alpha_{\mathrm{em}}^{2}\alpha_{s}^{2}$ & {\tt z2j} 
& $pp(\bar{p})$ $\rightarrow$ $Z/\gamma^{*}(2f)$ + 2 jets (+$X$) \\

113 & $\alpha_{\mathrm{em}}^{2}\alpha_{s}^{3}$ & {\tt z3j} 
& $pp(\bar{p})$ $\rightarrow$ $Z/\gamma^{*}(2f)$ + 3 jets (+$X$) \\

114 & $\alpha_{\mathrm{em}}^{2}\alpha_{s}^{4}$ & {\tt z4j} 
& $pp(\bar{p})$ $\rightarrow$ $Z/\gamma^{*}(2f)$ + 4 jets (+$X$) \\ \hline

120 & $\alpha_{\mathrm{em}}^{4}$ & {\tt ww0j} 
& $pp(\bar{p})$ $\rightarrow$ $W^{+}(2f)W^{-}(2f')$ + 0 jet (+$X$) \\

121 & $\alpha_{\mathrm{em}}^{4}\alpha_{s}$ & {\tt ww1j} 
& $pp(\bar{p})$ $\rightarrow$ $W^{+}(2f)W^{-}(2f')$ + 1 jet (+$X$) \\

122 & $\alpha_{\mathrm{em}}^{4}\alpha_{s}^{2}$ & {\tt ww2j} 
& $pp(\bar{p})$ $\rightarrow$ $W^{+}(2f)W^{-}(2f')$ + 2 jets (+$X$) \\ \hline

130 & $\alpha_{\mathrm{em}}^{4}$ & {\tt zw0j} 
& $pp(\bar{p})$ $\rightarrow$ $Z/\gamma^{*}(2f)W(2f')$ + 0 jet (+$X$) \\

131 & $\alpha_{\mathrm{em}}^{4}\alpha_{s}$ & {\tt zw1j} 
& $pp(\bar{p})$ $\rightarrow$ $Z/\gamma^{*}(2f)W(2f')$ + 1 jet (+$X$) \\

132 & $\alpha_{\mathrm{em}}^{4}\alpha_{s}^{2}$ & {\tt zw2j} 
& $pp(\bar{p})$ $\rightarrow$ $Z/\gamma^{*}(2f)W(2f')$ + 2 jets (+$X$) \\ \hline

140 & $\alpha_{\mathrm{em}}^{4}$ & {\tt zz0j} 
& $pp(\bar{p})$ $\rightarrow$ $Z/\gamma^{*}(2f)Z/\gamma^{*}(2f')$ + 0 jet (+$X$) \\

141 & $\alpha_{\mathrm{em}}^{4}\alpha_{s}$ & {\tt zz1j} 
& $pp(\bar{p})$ $\rightarrow$ $Z/\gamma^{*}(2f)Z/\gamma^{*}(2f')$ + 1 jet (+$X$) \\

142 & $\alpha_{\mathrm{em}}^{4}\alpha_{s}^{2}$ & {\tt zz2j} 
& $pp(\bar{p})$ $\rightarrow$ $Z/\gamma^{*}(2f)Z/\gamma^{*}(2f')$ + 2 jets (+$X$) \\ \hline

160 & $Y_{b}^{2}\alpha_{s}^{2}$ & {\tt b4hbb}
& $pp(\bar{p})$ $\rightarrow$ $h_{0}(b\bar{b})$ + $b\bar{b}$ (+$X$) \\

161 & $\alpha_{\mathrm{em}}^{2}\alpha_{s}^{2}$ & {\tt b4zbb}
& $pp(\bar{p})$ $\rightarrow$ $Z/\gamma^{*}(b\bar{b})$ + $b\bar{b}$ (+$X$) \\

162 & $\alpha_{s}^{4}$ & {\tt b4qcd}
& $pp(\bar{p})$ $\rightarrow$ $b\bar{b}b\bar{b}$ (+$X$) \\

163 & $Y_{b}^{2}\alpha_{\mathrm{em}}^{2}$ & {\tt b4hz}
& $pp(\bar{p})$ $\rightarrow$ $h_{0}(b\bar{b})$ + $Z/\gamma^{*}(b\bar{b})$ (+$X$) \\

164 & $\alpha_{\mathrm{em}}^{4}$ & {\tt b4zz}
& $pp(\bar{p})$ $\rightarrow$ $Z/\gamma^{*}(b\bar{b})$ + $Z/\gamma^{*}(b\bar{b})$ (+$X$) \\ \hline

170 & $\alpha_{\mathrm{em}}^{4}\alpha_{s}^{2}$ & {\tt tt6bdy}
& $pp(\bar{p})$ $\rightarrow$ $t\bar{t}$ $\rightarrow$ $6f$ (+$X$) \\

171 & $\alpha_{\mathrm{em}}^{4}\alpha_{s}^{3}$ & {\tt ttj7bdy}
& $pp(\bar{p})$ $\rightarrow$ $t\bar{t}$ + 1 jet $\rightarrow$ $6f$ + 1 jet (+$X$) \\ \hline

182 & $\alpha_{s}^{2}$ & {\tt qcd2j}
& $pp(\bar{p})$ $\rightarrow$ 2 jets (+$X$) \\

183 & $\alpha_{s}^{3}$ & {\tt qcd3j}
& $pp(\bar{p})$ $\rightarrow$ 3 jets (+$X$) \\

184 & $\alpha_{s}^{4}$ & {\tt qcd4j}
& $pp(\bar{p})$ $\rightarrow$ 4 jets (+$X$) \\ \hline

\end{tabular}
\end{center}
\caption{Processes included in GR@PPA, version 2.7.}
\label{tab:proc}
\end{table}

\section{Parameters}

The kinematical cuts and other parameters used in the benchmark tests in 
Section 4 are as follows:

\begin{list}{}{
\setlength{\leftmargin}{0.5cm}
\setlength{\labelwidth}{2cm}
\setlength{\labelsep}{-1.5cm}
}
\item[{\bf Beams} : $\quad$] $\;$ \\
The benchmark tests have been carried out for the Tevatron Run II ($p\bar{p}$) 
and the LHC ($pp$) conditions with the center-of-mass energies of
\begin{equation}
\sqrt{s} \; = \;
\Big{\{}
\begin{array}{clll}
1.96 & \mathrm{TeV} & p\bar{p} & \mathrm{(Tevatron \; Run \; II),} \\
14 & \mathrm{TeV} & pp & \mathrm{(LHC).}
\end{array}
\label{eq:beam}
\end{equation}

\item[{\bf Electroweak parameters} :] $\;$ \\
The $G_{\mu}$ scheme is adopted with the parameter set of 
\begin{equation}
(G_{F}, \; M_{W}, \; M_{Z}) \; = \; 
(1.16639 \times 10^{-5} \; \mathrm{GeV^{-2}}, \; 80.419 \; \mathrm{GeV}, \; 91.188 \; \mathrm{GeV}) \quad ,
\label{eq:gaugeinv}
\end{equation}
where $G_{F}$ is the Fermi constant, $M_{W}$ \cite{pdg2000} and 
$M_{Z}$ \cite{pdg2000} are the masses of the $W$ and $Z$ bosons, respectively. 
This set leads to the other parameters as
\begin{eqnarray}
	\alpha_{\mathrm{em}} \; = \; 1/132.51 \quad , \quad
	\sin^{2}\theta_{W} \; = \; 0.2222 \quad . \nonumber
\end{eqnarray}

\item[{\bf Boson width and decay} :] $\;$ \\
The $W$ boson is forced to decay to a pair of an electron and a neutrino, and 
the $Z$ boson to an electron-positron pair. In diboson production processes, 
the interference between electrons (neutrinos) from two bosons is ignored. We 
use fixed decay widths for $W$, $Z$ and the top quark as
\begin{equation}
	\Gamma_{W} \; = \; 2.048 \quad \mathrm{GeV}, \qquad
        \Gamma_{Z} \; = \; 2.446 \quad \mathrm{GeV}, \qquad
	\Gamma_{t} \; = \; 1.508 \quad \mathrm{GeV}.
\label{eq:fixedwidth}
\end{equation}

\item[{\bf CKM parameters} :] $\;$ \\
We use the following CKM parameters:
\begin{equation}
\left(
\begin{array}{c}
d' \\
s' \\
b'
\end{array}
\right)
\; = \;
\left(
\begin{array}{ccc}
|V_{ud}| = 0.9752 & |V_{us}| = 0.2210 & |V_{ub}| = 0.0054 \\
|V_{cd}| = 0.2210 & |V_{cs}| = 0.9743 & |V_{cb}| = 0.0419 \\
|V_{td}| = 0.0054 & |V_{ts}| = 0.0419 & |V_{tb}| = 0.9991
\end{array}
\right)
\left(
\begin{array}{c}
d \\
s \\
b
\end{array}
\right)
\quad .
\label{eq:ckmmatrix}
\end{equation}
The phase of CKM matrix is neglected.

\item[{\bf Fermion masses} :] $\;$ \\
Following values are use for the masses of heavy quarks 
\begin{equation}
m_{c} = 1.5 \; \mathrm{GeV} \quad 
m_{b} = 4.7 \; \mathrm{GeV} \quad
m_{t} = 174.3 \; \mathrm{GeV} \quad .
\label{eq:qmasses}
\end{equation}
The other quarks ($u$,$d$,$s$) are assumed to be massless. The electron and 
neutrino masses are also neglected.

\item[{\bf Proton Distribution Function} :] $\;$ \\
We use CTEQ6L \cite{cteqpdf} containing 5 flavors for PDF. The corresponding 
strong coupling constant is $\alpha_{s}(M_{Z}^2)$ = 0.1180.

\item[{\bf Energy scale} :] $\;$ \\
The renormalization and factorization scales ($Q$) are always chosen to be 
identical and fixed to the $Z$ boson mass as
\begin{equation}
Q \; = \; \mu_{F} \; = \; \mu_{R} \; = \; m_{Z} 
  \; = \; 91.188 \quad \mathrm{GeV.}
\label{eq:qsquare}
\end{equation}

\item[{\bf Kinematical cuts} :] $\;$ \\
The following kinematical cuts are applied to avoid infrared/collinear 
singularities,
\begin{equation}
p_{T} \; \geq \; 20 \; \mathrm{GeV} \; , \quad 
|\eta| \; \leq \; 3.0 \; , \quad 
\Delta R \; \geq \; 0.4 \quad .
\label{eq:kincuts}
\end{equation}
These cuts are applied to the jets ($u$, $d$, $c$, $s$, $b$ and $g$) and 
electrons, while no cut is done to top quarks and neutrinos.

\item[{\bf Flavor selection} :] $\;$ \\
In some tests of heavy flavor productions, a particular jet flavor is 
selected in the integration. This is done in the subroutine {\tt GRCUSRCUT}. 
The following is an example to select those events having two top quarks in 
the final state:
\renewcommand{\baselinestretch}{0.78}
\begin{verbatim}
   ITOP = 0               ! counter of the number of top-quark in jets.
   DO I = 1,NJET          ! loop over n jets.
      IF (IABS(GRCKFCD(I)).EQ.6) ITOP = ITOP + 1  ! number of top qaurks.
   ENDDO
   IF (ITOP.NE.2) IUSRCUT = 1   ! reject this event.
\end{verbatim}
\ \\
\end{list}

\end{document}